\documentclass[aip,rsi,amsmath,amssymb,reprint,floatfix]{revtex4-1}
\pdfoutput=1
\usepackage[totalwidth=480pt, totalheight=660pt]{geometry}
\usepackage{graphicx}
\usepackage{dcolumn}
\usepackage{bm}
\usepackage{hyperref}
\usepackage{subfig}
\usepackage{caption}
\usepackage[export]{adjustbox}
\begin{document}
\preprint{AIP/123-QED}

\title{Energy Transfer Times in Fluorographene-based Biomimetic Light Harvesting Antennae}

\author{Sayeh Rajabi}
\author{Tom\'{a}\v{s} Man\v{c}al}
 \email{mancal@karlov.mff.cuni.cz}
\affiliation{ 
Faculty of Mathematics and Physics, Charles University,\\ Ke Karlovu 5, 121 16 Prague 2, Czech Republic
}

\begin{abstract}
It has been demonstrated earlier that graphene-like defects on fluorographene can act as molecules in biomimetic molecular light-harvesting antennae. In competition with radiative and non-radiative losses, transfer time of excitations in an antenna measures its performance. We report on the optimal conditions for excitation energy transfer in artificial antennae built from selected types of fluorographene defects. The excitation transfer dynamics is calculated based on the Frenkel exciton model using hierarchical equations of motion for different values of temperature, system-environment reorganization energy and bath correlation time to study a possible range of parameters pertaining to the fluorographene material.  We also study possible energy funnelling in the third dimension for two parallel fluorographene sheets with defects. We conclude that the strength of system-environment interaction is more important for the efficient energy funnelling in our proposed material than a precise control over the structure of artificial antennae.

\end{abstract}

\maketitle

\section{Introduction}

Natural photosynthetic light-harvesting systems remain an inspiration
for research into solar energy converting materials. Despite the
relatively low overall efficiency of light to usable chemical energy conversion
exhibited by plants and bacteria \cite{blankenship2011}, the initial steps from light absorption to charge separation, are characterized by a remarkable, close
to unity, quantum efficiency. Mimicking this efficient initial process,
and connecting it to some artificial machinery, which would keep the overall process efficient, remain one of the promising strategies 
of the research into biomimetic light-harvesting systems \cite{scholes2011}. Photosynthetic light-harvesting systems are highly organized collections of basic building blocks, pigment molecules, embedded in protein matrix. The protein environment both ensures the organization and induces various types of disorder and fluctuations of the pigment properties on a wide range of time scales \cite{Maly2016c}. Theory of photosynthetic light-harvesting strongly benefits from the concept of Frenkel exciton \cite{van2000photosynthetic}, which is at the core of our present understanding of this remarkable set of processes. Photosynthetic systems, on the other hand, provide an excellent playground for development of theoretical methods for description of the dynamics of Frenkel exciton systems, as they do not straightforwardly yield to perturbation theory\cite{ishizaki2009a}. Theoretical difficulties in correctly describing excited state dynamics of such systems have stimulated development of exact methodologies in recent years \cite{ishizaki2009unified}. A generation of biologists, chemists and physicists, of which a remarkable number celebrates seventieth birthday this year (49ers \cite{fleming2018}), has been largely responsible for disentangling the intricacies of the natural light-harvesting and primary processes of photosynthesis in general. This work is a part of the Festschrift dedicated to Leonas Valkunas, one of the 49ers whose contributions to the theory of photosynthetic excitons still keep the field moving forward \cite{van2000photosynthetic, valkunas2013}. 

In nature, photosynthesis is always a process occurring on membranes \cite{ruban2013a, blankenship2014molecular}, quasi two-dimensional structures dividing the aqueous cell environment into two parts. The photosynthetic apparatus, residing in the membrane, facilitates transfer of charged particles between the two environments. To construct such a membrane-based artificial light-harvesting antenna would be a straightforward generalization of the main design principle of natural photosynthetic light-harvesting. The omnipresence of graphene in recent discussions of nanotechnology makes considering it in place of a photosynthetic membrane inevitable. Two-dimensional materials are in fact subject to great surge in research activity \cite{briggs2019roadmap, das2019role,hao2019graphane}, with many new materials realized, including derivatives of graphene \cite{nair2010fluorographene}, one of which is Fluorographene (FG) \cite{zbovril2010graphene, nair2010fluorographene}.

The idea that graphene-like impurities in FG lattice  can be treated as molecules, when excited by light, was introduced in our
previous work\cite{slama2018fluorographene}. Graphene-like impurities or defects in FG are compact regions of the FG sheet, where fluorine atoms have been removed from the lattice, forming isles of conjugated bonds. These isles of different shapes and sizes are envisioned to play the role of chromophores similarly to (bacterio)chlorophyll in photosynthetic light-harvesting systems of plants and bacteria. The molecular approach we proposed for the calculations of excitation energy transfer is in turn inspired by the well-tested Frenkel exciton methodology used in the photosynthetic light-harvesting community \cite{may2008charge,van2000photosynthetic}. This approach can be in principle  applied to arbitrarily large aggregates of random defects on a single layer as well as to a stack of such FG layers. Especially when the whole extended system of light-harvesting antennae is composed of subsystems which are mutually weakly coupled (such as small groups of strongly coupled defects) typical traditional solid-state techniques assuming periodic structure would be less suitable for studying the properties of extended arrangements of defects. For instance, if the coupling between moleculoids---defects behaving like molecules---residing on different sheets could be deemed weak, energy transfer between sheets could be treated by simple rate theories and the size of the problem increases only linearly with the number of sheets (no diagonalization of the multi-sheet Hamiltonian would be needed). 

The main findings of our previous work\cite{slama2018fluorographene} are that (1) perylene- and anthanthrene-like impurities on the FG lattice are stable and their molecular orbitals, as well as transition dipole moments, are well localized over the region of the defects, and (2) energy funnelling in sample antennae made of such defects positioned in locations constrained by the FG lattice achieves sufficiently short energy transfer times to out-compete expected 
loss channels. These results are paving the way for constructing artificial light-harvesting systems based on FG and possibly on other graphene derivatives, or similar two-dimensional materials. 

The present work centers around analysis of antenna geometries towards generalizing the previous results to include random antenna configurations and towards finding optimal conditions for the excitation energy transfer. To the best of our knowledge, virtually nothing
is known about the control of the particular defects in FG sheet which we study. It is reasonable to assume that if the material
we propose were to be synthesized, there would be a limited control over the precise arrangement of the defects and possibly also over the
orientation and size of the defects. In order to mimic the variability of the antennae and to estimate how much control would be needed
to achieve the desired function, we assume the defects in a FG lattice to occur with random spatial distribution. We limit our study to two types of defects for which we have determined the most important properties\cite{slama2018fluorographene}. In particular, we assume the perylene-like and anthanthrene-like defects, which correspond to the smallest molecules built out of benzene rings (the pattern which repeats in the graphene) whose first excitation energies lie within the energy gap of FG, that is above 3 eV. If the defects are distant enough (separation by one row of fluorine seems to be sufficient \cite{slama2018fluorographene}), we can neglect their mutual orbital overlap and hence electron transfer between them. The system thus fulfills the conditions
for quantitative application of the Frenkel exciton model \cite{seibt2016optical} to predict the electronic state structure of the 
aggregate of defects. Our aim is to study the possible process of energy concentration on low energy acceptor moleculoids occurring either in the same FG sheet as the donor (artificial light-harvesting antenna) or on a different sheet. In both cases we assume that further components of the light-harvesting machinery would have to be added to complete the light energy conversion. 

The paper is organized as follows: In Section \ref{Model} we introduce the model of an artificial light-harvesting antenna based on graphene-like defects in a fluorographene sheet. We discuss randomized structures based on this model and the simulations of the excited state dynamics, in particular energy transfer time between model antennae and an acceptor molecule, in Section \ref{simulations}. Properties of the structures occurring on a single FG sheet are discussed in Section \ref{dis_2D}, while in Section \ref{dis_3D} we focus on a limited number of arrangements involving two FG sheets. We conclude and provide an outlook in Section \ref{concl}.

\section{Model of the Light-Harvesting Antenna}
\label{Model}
%
%
\begin{figure}[t]
  \includegraphics[width=\columnwidth]{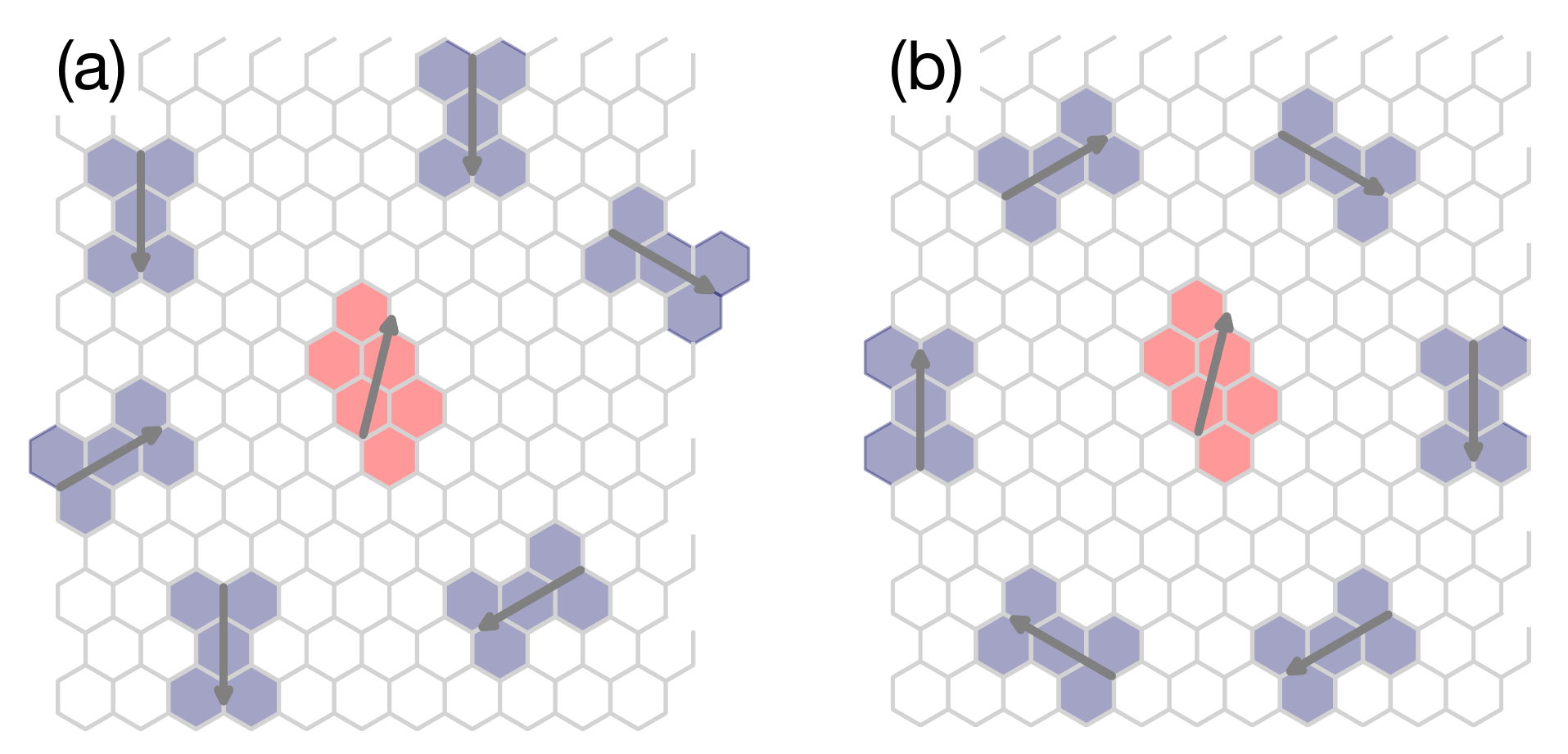}
  \caption{\label{fig1} (a) Random and (b) symmetric antennae on the antenna cell, each made out of six perylene-like defects around one anthanthrene-like defect located at the center. Arrows show the direction of transition dipole moments.}
\end{figure}
%
%
%

%
%
\begin{figure}
  \includegraphics[width=\columnwidth]{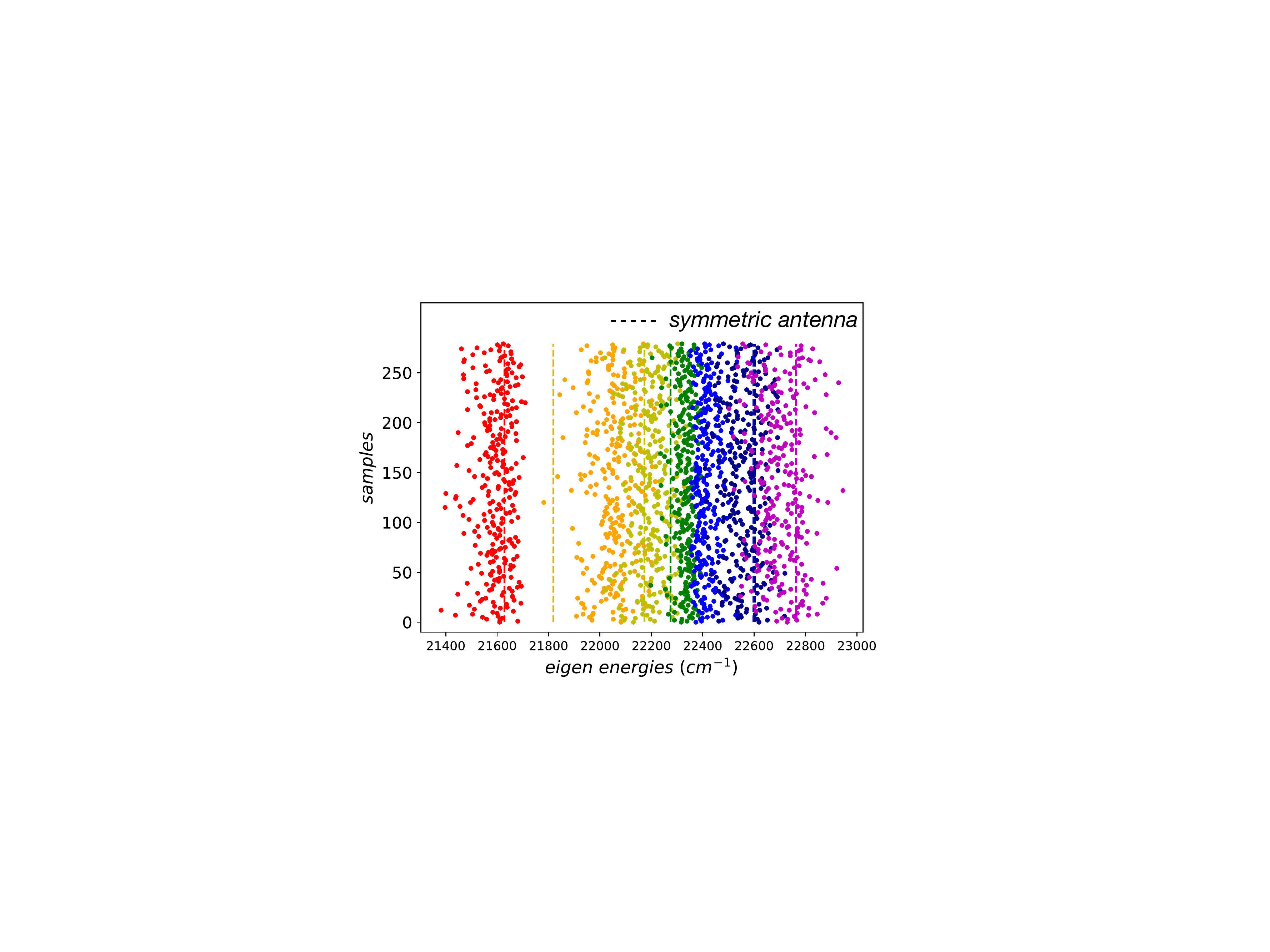}
  \caption{\label{fig2:eig_dist}\small Distribution of the eigenvalues of 280 system Hamiltonians studied in this work. Colors represent different eigenvalues. Note also that the first excitation energies of perylene- and anthanthrene-like defects are $E_{P}=22354 \ \text{cm}^{-1}$ and $E_{A}=21736 \ \text{cm}^{-1}$ respectively. Dashed lines represent the eigen-energies of the symmetric antenna, Fig. \ref{fig1}b.}
\end{figure}

Light-harvesting antennae occurring in nature can be efficiently treated as groups of two-level systems interacting through electrostatic resonance coupling \cite{van2000photosynthetic}. It seems to be a specific design principle of these antennae that the molecular orbitals of the neighboring molecules do not overlap, ensuring that only excitation energy, and not charged particles, are transferred during the initial excitation energy transfer. For the theory this means a great deal of simplification in electronic structure of these systems \cite{seibt2016optical}. In our artificial antenna, we assume the same principle to be realized. Each moleculoid is approximated by its ground and excited states, $|g\rangle$ and $|e\rangle$ respectively, and hence the whole aggregate of moleculoids is described in terms of the coupled two-level systems. In the basis of the collective states $|i\rangle = |g_1\rangle \cdots |e_i\rangle \cdots |g_N\rangle$, where excitation is localized at moleculoid $i$ (further referred to as site $i$), the electronic Frenkel Hamiltonian of an antenna reads
\begin{align}
H_e = \sum\limits_{i} E_i |i\rangle\langle i | + \sum\limits_{i\neq j} J_{ij} |i\rangle\langle j |.
\label{eq:Frenkel}
\end{align}
Here, $E_i$ is the excitation energy of the $i$th moleculoid and $J_{ij}$ is the coupling between excited states at sites $i$ and $j$. The couplings are assumed to correspond to Coulombic dipole-dipole interaction of the transition dipole moments, $\vec{d}_i$, and they are given by
\begin{align}
J_{ij} = \frac{\kappa}{4 \pi\epsilon_0}\frac{\vec{d}_i \cdot \vec{d}_j - 3 (\vec{d}_i \cdot \hat{r}) (\vec{d}_j \cdot \hat{r})}{r^3},
\label{eq:dipole-dipole}
\end{align}
where $\hat{r}$ is the unit vector along $\vec{r}_{ij}=\vec{r}_i - \vec{r}_j$ and $r = |\vec{r}_{ij}|$. We have assumed a scaling factor $\kappa = 1.43$, which includes the effect of the environment, and scales the dipole-dipole approximation for the resonance couplings to agree with the values computed by quantum chemistry and reported in our previous work\cite{slama2018fluorographene}. Equivalently, we can assume that the effective relative permittivity of FG is $\epsilon_r^{eff}=0.7$. The fact that the effective relative permittivity is smaller than one is extensively discussed and justified\cite{slama2}, where an efficient method for calculation of couplings between moleculoids in FG sheet, including the effect of the sheet on the coupling, is studied. 

Electronic states of the system of moleculoids interacting with the FG lattice represent an open quantum system interacting with its environment. The dynamics of excitation in this system can be studied in different regimes of system-environment coupling strength (as measured by the reorganization energy) either by approximate methods, such as F\"orster and Redfield theories\cite{may2008charge, van2000photosynthetic}, or by numerically exact methods which became available during the past decade\cite{tanimura1989time, ishizaki2009unified, tanimura2012reduced, Oviedo-Casado2016, Rosenbach2016}. Because it is not known how strong the system-environment interaction in real FG systems is, we will investigate a range of interaction strengths. One of the main problems of approximate methods is their inability to interpolate between different system-environment interaction strength regimes. For this reason we choose to apply a formally exact approach, hierarchical equations of motion (HEOM) \cite{tanimura1989time, ishizaki2009unified, tanimura2012reduced}, for the treatment of excitation transfer in our artificial antennae. The requirement to use an exact propagation method such as HEOM, and a simultaneous desire to investigate a relatively large set of sample structures, limit our investigation to relatively small antennae. Our choice is therefore to take inspiration from naturally occurring bacterial antennae, the LH2 and LH1 systems of purple bacteria\cite{McDermott1995,Hu1998}, and to construct a smaller artificial antenna which keeps their main characteristics, namely their high circular symmetry. We choose to start with a symmetric artificial structure, and to investigate a randomized version of this artificial antenna. 

The task of the antenna is to capture light energy (a process we do not simulate in this work), and to deliver the energy to an acceptor at lower energy. Our antennae are constructed of perylene-like moleculoids, while the acceptor is (usually) represented by a single anthanthrene-like moleculoid. Locations and orientations of defects (moleculoids) and their transition energies are the system variables that, together with the environmental parameters (reorganization energy, correlation time and temperature), determine the performance of the antennae. In principle, the excitation transfer can also occur in three dimensions between defects in FG sheets. For demonstration, we also consider two parallel FG sheets with defects arranged in both symmetric and random structures.

When we describe the energy levels of the system in the sections that follow, we discuss the electronic excited states only. The term lowest energy level therefore refers to the lowest excited energy level.

\section{Simulations}
\label{simulations}

%
%
\begin{figure}[t]
  \includegraphics[width=0.48\textwidth]{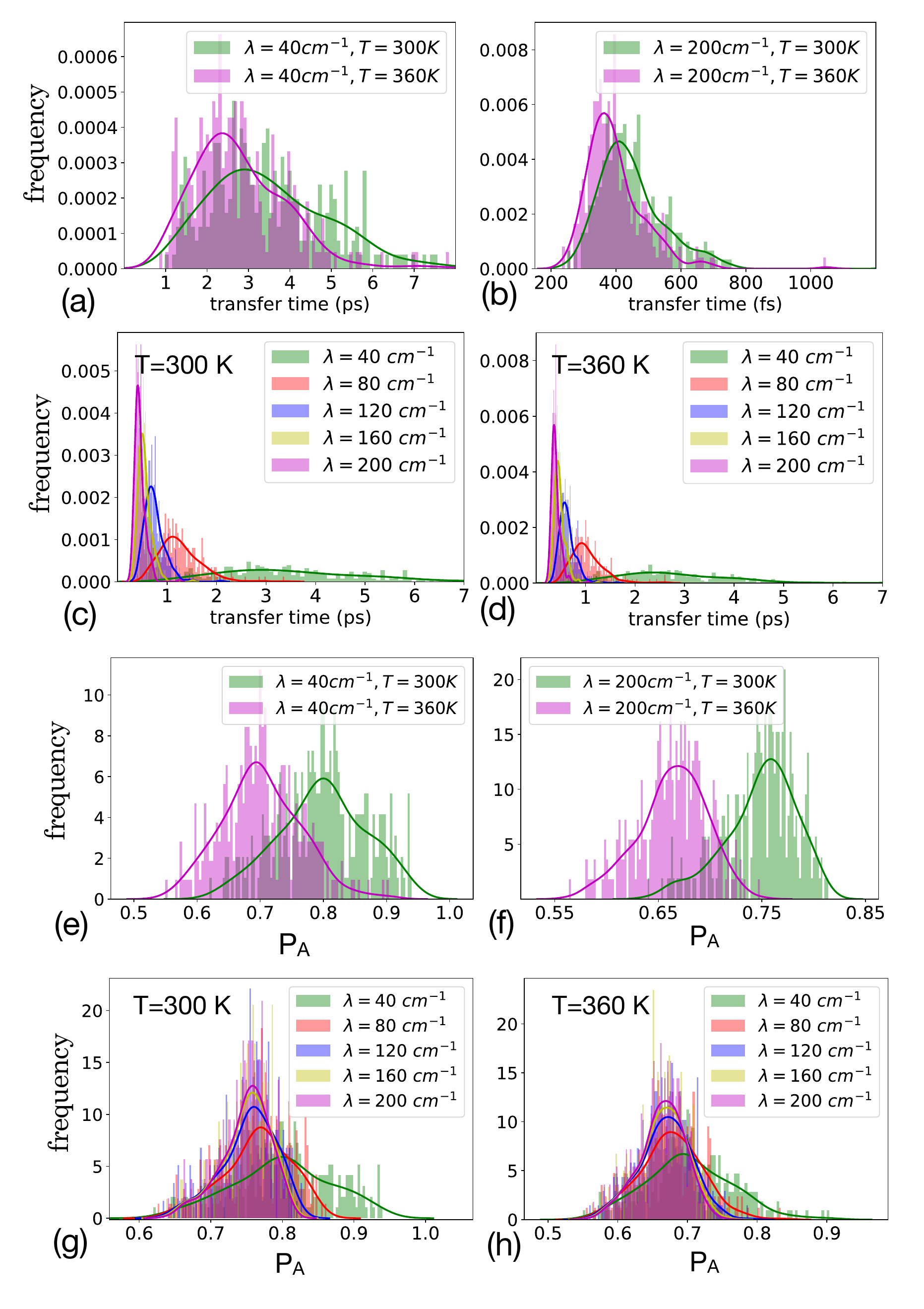}
  \caption{\label{fig3}Distribution of transfer times and final populations on the acceptor in all 280 random samples generated for this study for different values of reorganization energy $\lambda$ and temperature $T$. Shorter $\tau_A$ and slightly lower $P_A$ correspond to higher system-environment coupling and higher temperature (T = 360 K). Frequency axis in (a)--(d) is scaled such that the area under each curve is unity.}
\end{figure}
%
%
%

%
%
\begin{table}[t]{}
\scalebox{0.82}{
\begin{tabular}{c|c|c|c|c|c||c|c|c|c|c|}
	\cline{2-11}
	\multicolumn{1}{l|}{}           & \multicolumn{5}{c||}{$T=300$ K} & \multicolumn{5}{c|}{$T=360$ K} \\ \hline
	\multicolumn{1}{|c|}{$\lambda \ (\text{cm}^{-1})$} & 40  & 80 & 120 & 160 & 200 & 40  & 80 & 120 & 160 & 200 \\ \hline
	\multicolumn{1}{|c|}{$\langle \tau_A \rangle (\text{fs})$} & 3512 & 1259 & 741 & 544 & 450 & 2787 & 1031 & 631 & 475 & 398 \\ \hline
	\multicolumn{1}{|c|}{$\sigma_{\tau} (\text{fs})$}  & 1438 & 420 & 209 & 137 & 108 & 1114 & 35 & 170 & 114 & 92 \\ \hline
	\multicolumn{1}{|c|}{$\langle P_A \rangle$} & 0.799 & 0.764 & 0.753 & 0.749 & 0.749 & 0.702 & 0.676 & 0.668 & 0.665 & 0.664  \\ \hline
	\multicolumn{1}{|c|}{$\sigma_{P_A}$} & 0.069 & 0.048 & 0.039 & 0.036 & 0.034 & 0.062  & 0.045 &0.038  & 0.035 & 0.033  \\ \hline \hline
	\multicolumn{1}{|c|}{$\tau_A^{\text{sym}} (\text{fs})$} & 1260& 605& 423& 345 & 305 & 1056 & 520 & 371 & 306 & 274  \\ \hline
	\multicolumn{1}{|c|}{$P_A^{\text{sym}}$} & 0.608 & 0.611 & 0.617 & 0.624 & 0.630 & 0.543 & 0.545 & 0.549 & 0.554 & 0.558  \\ \hline
\end{tabular}}
\caption{\label{tab:table-meanstd}\small Mean and standard deviation values for transfer time and final populations on acceptors in 280 random samples, with fixed bath correlation time $\gamma^{-1}=60$ fs. $\tau_A^{\text{sym}}$ and $P_A^{\text{sym}}$ correspond to the symmetric antenna.}
\end{table}  
%
%
%

%
%
\begin{figure}
  \centering
  \includegraphics[width=0.48\textwidth]{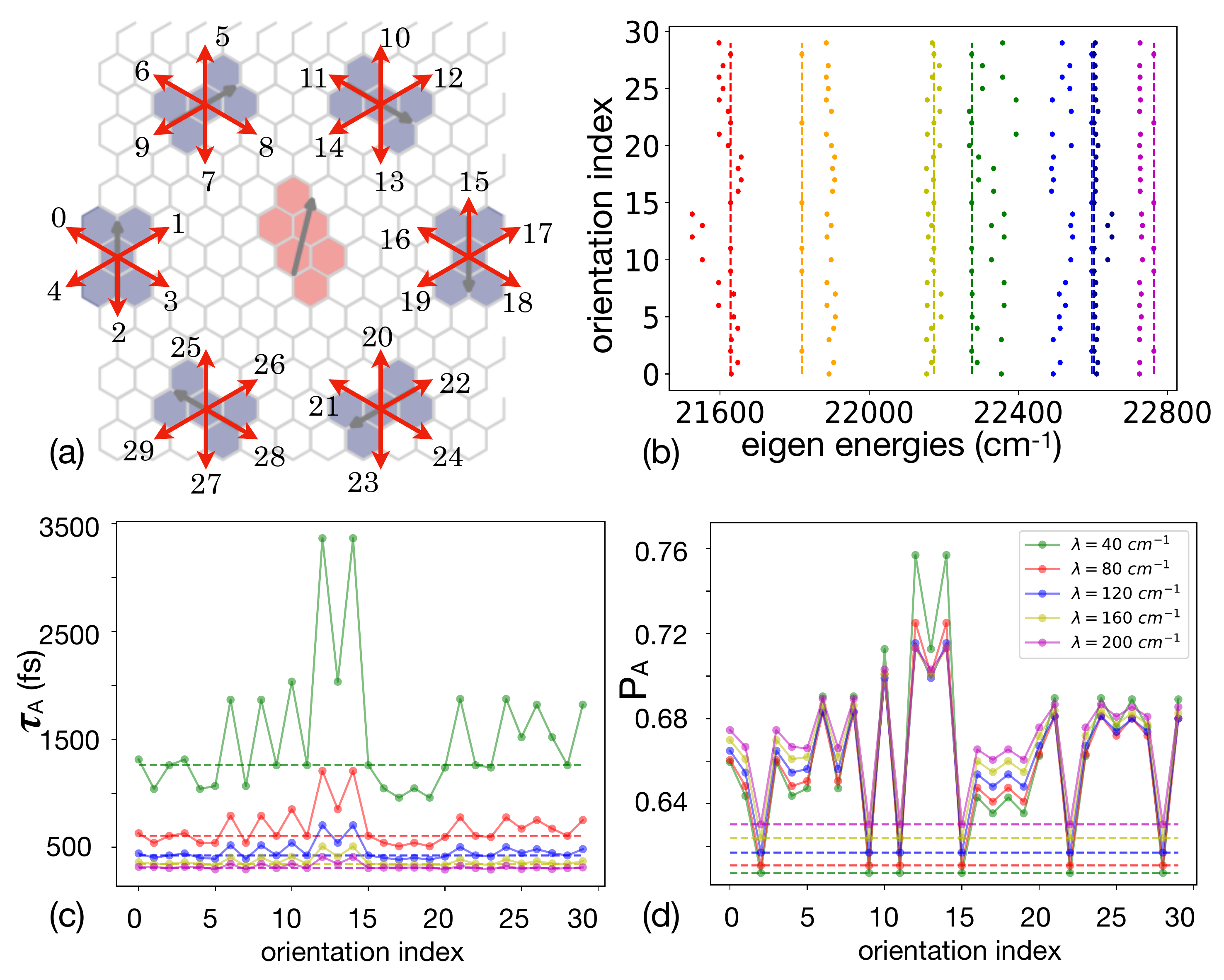}
  \caption{\label{fig4}\small Symmetric antenna (see Fig. \ref{fig1}b) under variations of dipole moment orientation. We have assumed that only one moleculoid is rotated at a time and others are kept along their original direction (un-indexed arrows in (a)). (b) Distribution of eigen-energies of the corresponding system Hamiltonians. (c) Transfer time to the acceptor and (d) final population of the acceptor for various orientations of perylene moleculoids. Dashed lines in (b), (c) and (d) correspond to the symmetric geometry.}
\end{figure}

We study 280 random samples of six perylene-like defects (referred to as `perylenes' from now on) with first excitation energy $E_P=22354 \ \text{cm}^{-1}$, randomly placed around an anthanthrene with $E_A=21736 \ \text{cm}^{-1}$ located in the middle of an \textit{antenna cell}, all in a single FG lattice. These values of transition energy and the ones for transition dipole moments, $d_P = 6.30\ D$ and $d_A = 6.55\ D$, are obtained from quantum chemistry calculations \cite{slama2018fluorographene} of a moleculoid placed in the FG sheet. An antenna cell is considered to be an area of $11 \times 14$ lattice cells ($\simeq 21.00 \times 28.58 \ \text{\AA}^2$ with lattice constant $a=1.5$ \AA). The location of anthanthrene is assumed to be fixed and only perylenes are distributed randomly. We also study a symmetric antenna in which the perylenes form a hexagonal ring around the anthanthrene. This shape is inspired by the structure of the LH2 antenna of purple bacteria, but the number of participating moleculoids is limited to six. Note that we limit the dimension of the sheet only for the purpose of the placement of the defects. The results of our calculations should be considered valid for a (possibly infinite) sheet in which all defects are placed sufficiently far away from the sheet edges (and each other), so that they do not influence the properties of the defects. These were, in fact, the conditions we used for calculation of the defect properties by quantum chemistry\cite{slama2018fluorographene}.

Fig. \ref{fig1} shows both a randomized and the symmetric antennae. Notice that here the `symmetric' geometry only refers to the positions and orientations of the perylene moleculoids, as the geometry of anthanthrene always breaks some symmetry on the hexagonal lattice. The algorithm we follow to generate antenna samples considers a two-lattice-cell distance between the areas occupied by adjacent moleculoids to guarantee zero orbital overlap between them. Generated samples (sets of transition dipole moments and positions of the center of moleculoids) are processed by the Quantarhei package \cite{quantarhei} to compute and diagonalize their Frenkel exciton Hamiltonians, Eq. (\ref{eq:Frenkel}), based on dipole-dipole approximation for the interaction between moleculoids. Fig. \ref{fig2:eig_dist} presents  eigenvalues of all sample Hamiltonians for which excited state dynamics is calculated and analyzed in this work.

In the absence of experimental data, we assume that the FG environment is characterized by over-damped Brownian oscillator spectral density $J(\omega)=\frac{2\lambda\gamma\omega}{\omega^2+\gamma^2}$, a model with minimal assumptions, characterized by two parameters, the reorganization energy $\lambda$ and correlation time $\tau_c=1/\gamma$. Parameters in the simulations include the reorganization energies $\lambda =$ 40,  80, 120, 160, 200 $\text{cm}^{-1}$, correlation times $\gamma^{-1}=$ 40, 60, 80, 100, 120, 140, 160, 180, 200 fs, and two temperatures $T = 300\ \text{and}\ 360 \ K$. The simulation time step is 1 fs and the runtime is 70 ps for our smallest reorganization energy $\lambda=40\ \text{cm}^{-1}$, and 30 ps for other values of $\lambda$ to ensure reaching the equilibrium in every sample and for every set of parameters.

Initial population is equally distributed over the perylenes (donors); and anthanthrene (the acceptor) has initially zero population. The time-evolution of the density matrix is calculated using HEOM, implemented in the QMaster package \cite{kreisbeck2014scalable,  kreisbeck2012long, kreisbeck2011high}, and is limited to the truncation level 4 of Kubo-Tannimura hierarchy. We verified that this level of hierarchy leads to convergency in the population dynamics. Populations in the moleculoid sites are the output data of the simulations. Although the eigenstates of the system do not in general coincide with the states localized on different molecules, due to the relatively large energy gap between anthanthrene- and perylene-like moleculoids, the lowest eigenstate level largely coincides with the anthanthrene state. 

For all samples with different parameters considered, energy funnelling towards the acceptor (anthanthrene) occurs and a single-exponential curve $p_A(t) = P_A (1-e^{-t/\tau_A})$ approximates the evolution of acceptor's population to its maximum value $P_A$ well. 
This behaviour is in fact expected as the energy gap between donor (antenna) and the acceptor is relatively large in all configurations, and it is expected to form the bottle neck of the transfer process. Note that the final step of energy transfer between the LH1 system and the reaction center, also forms such a bottleneck in naturally occurring systems \cite{blankenship2014molecular}. We use the least-square fitting method to obtain the final population of the acceptor $P_A$ and transfer time $\tau_A$, the quantities that in principle characterize the performance of an antenna. The crucial quantity in this study is the transfer time. By connecting the acceptor to a sink (i.e. a further step of energy conversion process, such as charge generation, or another antenna on lower energy), one can continuously extract the accumulated excitations from the acceptor. We assume excited state life-times on the order of nanoseconds, as in photosynthetic systems. The ratio between the transfer time and the expected excited state life-time of the moleculoids corresponds to the ratio of lost excitations. For transfer time faster that 100 ps and life-time on the order of nanoseconds, the quantum efficiency of the transfer process is always above 90~\%.  As we are only interested in comparing antennae between each other, we simply assume that the faster is the population transfer to the acceptor, the better is the performance of the antenna.

\section{Single Sheet Antennae}
\label{dis_2D}

%
%
\begin{figure}
  \centering
  \includegraphics[width=\columnwidth]{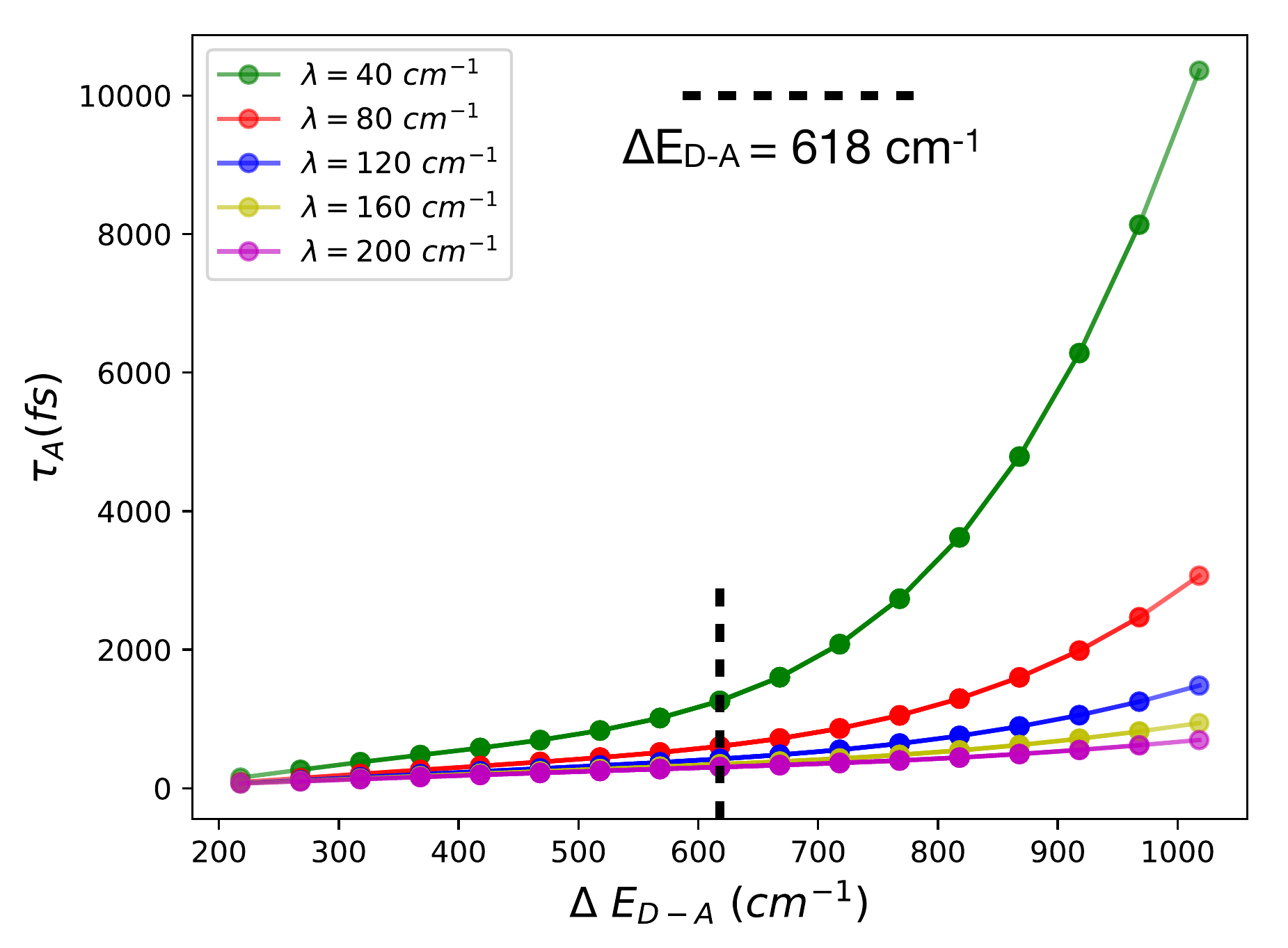}
  \caption{\label{fig5}\small Transfer times in the symmetric antenna (see Fig. \ref{fig1}b) for various energy gaps $\Delta E_{D-A}$ between the excited states of the acceptor and donor moleculoids. For perylene- and anthanthrene-like moleculoids considered in this work, $\Delta E_{D-A} = 618\ \text{cm}^{-1}$. For all values of $\lambda$, energy transfer is always faster with smaller energy gaps.}
\end{figure}
%
%
%

%
%
\begin{figure}
  \centering
  \includegraphics[width=\columnwidth]{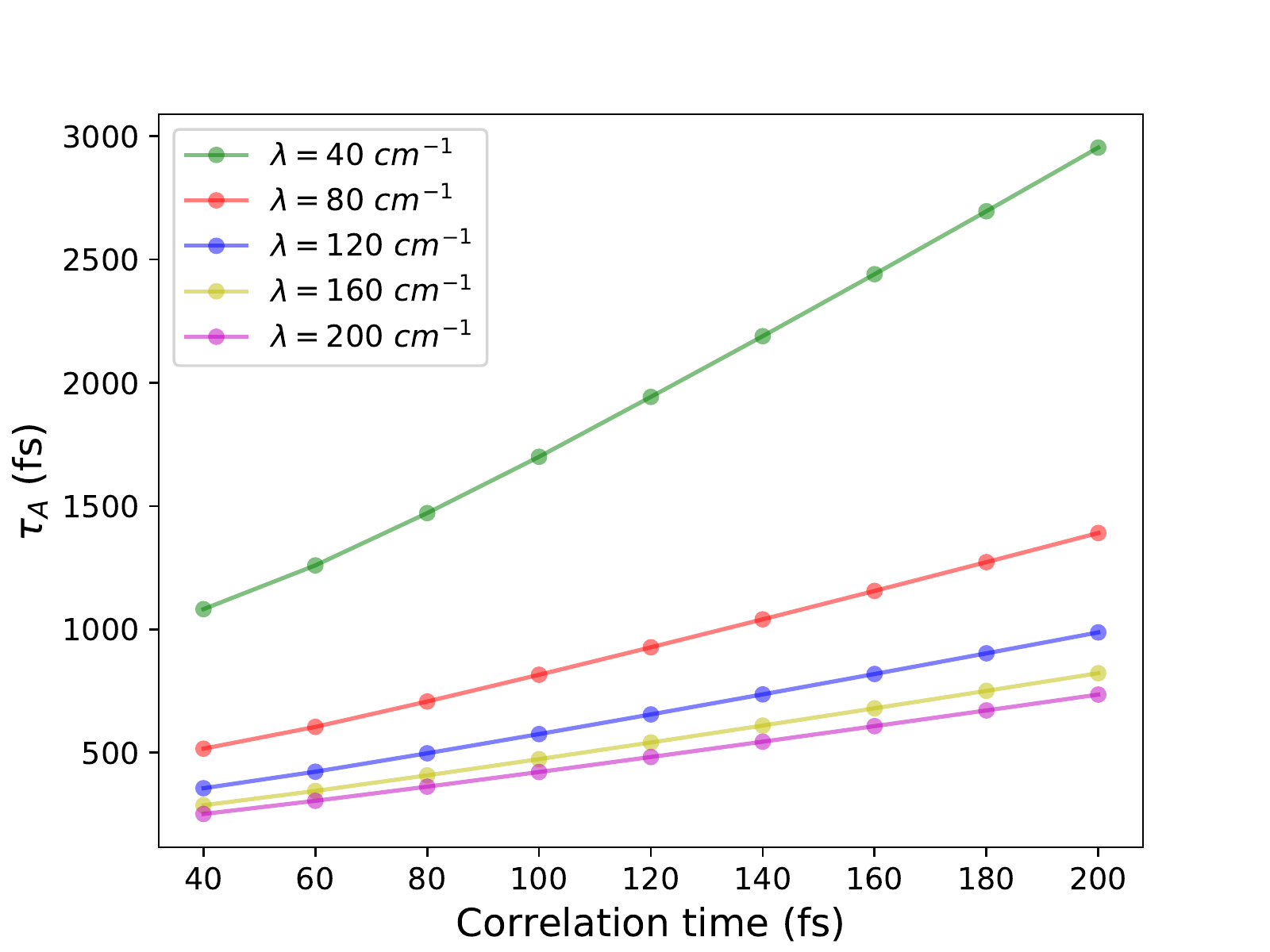}
  \caption{\label{fig6}Transfer time in the symmetric antenna as a function of the correlation time for different values of reorganization energy at $T=300$ K.}
\end{figure}

Distribution of transfer times in random samples on a single FG sheet calculated for different values of reorganization energy and temperature is shown in Figs. \ref{fig3}a--d, where we have kept fixed the bath correlation time $\gamma^{-1}=60$ fs. The plots show histograms and fitted normalized (Gaussian) kernel density estimate (KDE) with bandwidth 3.\footnote{KDE sums over normal Gaussian curves drawn centered at each observation. The bandwidth of KDE---similar to bin size in histograms---corresponds to the width of the kernels.}
Similarly we plot the distribution of final populations on acceptors, Figs. \ref{fig3}e--h. It is clear from the plots that shorter transfer times---faster excitation transfer and energy funnelling---correspond to larger reorganization energies and the higher temperature ($T=360$ K). Exciton transfer in the studied samples occurs as fast as in few hundred femtoseconds for large enough reorganization energies. Fig. \ref{fig3} (also see SI) clearly show diversity (relatively large range) in the values of $P_A$ and $\tau_A$. This diversity is a direct consequence of the variations of the geometry and hence the electronic structure of the antennae for fixed environmental parameters. Note also that higher diversity (broader histograms) in both $\tau_A$ and $P_A$ correspond to lower values of reorganization energy, while the major effect of $\lambda$ is on the diversity of $\tau_A$. Hence, geometric optimization for fast excitation transfer in the antennae mainly matters when the system-environment coupling is weak and transfer time has a wide range of values. The mean and standard deviation values for $\tau_A$ and $P_A$ of the random samples are given in Table \ref{tab:table-meanstd} which also contains the transfer time and final population on the acceptor in the symmetric antenna.

Our calculations show that $\tau_A$ always decreases as $\lambda$ increases with the environmental parameters we study (see SI). Earlier we also studied situations in which increasing the reorganization energy leads to saturation or even to a small rise of transfer times\cite{slama2018fluorographene}. These effects however occur in a regime in which the transfer is nevertheless very fast. The change of $P_A$ with $\lambda$ is negligible in most geometries and more noticeable for higher values of $P_A$ (which itself is determined by larger energy gap). As $\lambda$ increases, higher values of $P_A$ slowly decrease, but for lower values of $P_A$, the change is either negligible or they slightly increase. We also conclude that, for all reorganization energies and both temperatures, the symmetric geometry, Fig. \ref{fig1}b, results in considerably shorter transfer time compared to average random samples, hence is one of the best performing structures, and at the same time it gains the lowest population on its acceptor. This reflects the fact that the symmetric antenna has its lowest energy level closer to the acceptor level than most of the random antennae.

To further examine the symmetric antenna and particularly its robustness, we consider some discrete fluctuations around the symmetric geometry. We first keep the positions of all monomers fixed and vary the orientations of perylene moleculoids, one at a time, Fig. \ref{fig4}. There are hence 30 configurations for such antennae. Note also that $\theta \rightarrow \theta + \pi$ transformation does not change the dynamics and hence the transfer time in the antenna. For our particular parameters, $T=300$ K and $\gamma^{-1}=60$ fs, the transfer time $\tau_A$ varies between few hundreds of femtoseconds for $\lambda=200$ cm$^{-1}$ to almost 3.5 ps for $\lambda=40$ cm$^{-1}$ and unfavorable geometries. Figs. \ref{fig4}b--c show that the slow transfer coincides with large energy gap between the lowest and the second lowest energy level of the system and correspondingly also with the high final population of the acceptor. As is expected, the variations are larger when the reorganization energy is smaller and resonance couplings, Eq. \ref{eq:dipole-dipole}, are the dominant factor in the exciton dynamics. For examples, when $\lambda = 40\ \text{cm}^{-1}$, exciton transfer in orientational configuration 12 occurs about three times faster than in 16.  

Resonance couplings between the moleculoids are also distance dependent. We changed the location of moleculoids, one at a time by one lattice-site, assuming the required distance between them to avoid orbital overlap, while keeping the dipole orientations fixed as in the symmetric geometry. This in total leads to 22 configurations (see SI). Comparison of the changes in $\tau_A$ due to orientation with those due to perylene-anthanthrene distance show more significant fluctuations when the orientations are varied.

The energy gap between perylene and anthathrene is $E_P - E_A = 618\ \text{cm}^{-1}$, as calculated by quantum chemistry methods. Changes in the transfer time when the energy gap varies are shown in Fig. \ref{fig5} for the symmetric antenna with different values of the reorganization energy. We have assumed $T=300$ K, $\gamma^{-1}=60$ fs and the energies varying between $21736 \pm 200\ \text{and}\ 22345 \pm 200 \ \text{cm}^{-1}$ with $50 \ \text{cm}^{-1}$ steps. The most significant observed changes correspond to larger energy gaps and smaller reorganization energy. We conclude that for moleculoids with smaller energy gaps between their first excited states, excitation transfer is faster. If we project this result on the problem involving general antennae, realizing that the changes of the donor-acceptor energy gap seem to be the main factor of the geometric optimization, we can also conclude that for large reorganization energies geometric optimization cannot enhance the excitation transfer significantly, as $\tau_A$ has a negligible dependence on the energy gap.

Finally, we study the effect of correlation time, another environmental parameter, on the transfer times. The correlation time influences the shape of the environment spectral density, and therefore modulates the energy transfer rates. For the parameters of our artificial antennae, longer environment correlation times lead to longer transfer times. In Fig. \ref{fig6} we show that in the symmetric antenna, transfer time almost linearly increases with the correlation time for all values of $\lambda$ that we investigated, and fixed temperature $T=300$ K. See SI for dependence of $P_A$ on the correlation time, as well as a more detailed parameter space of random antennae formed by selected values of both reorganization energy and correlation time.

\section{Antennae with Two Sheets}
\label{dis_3D}
%
%
%
\begin{figure}
  \includegraphics[width=0.36\textwidth]{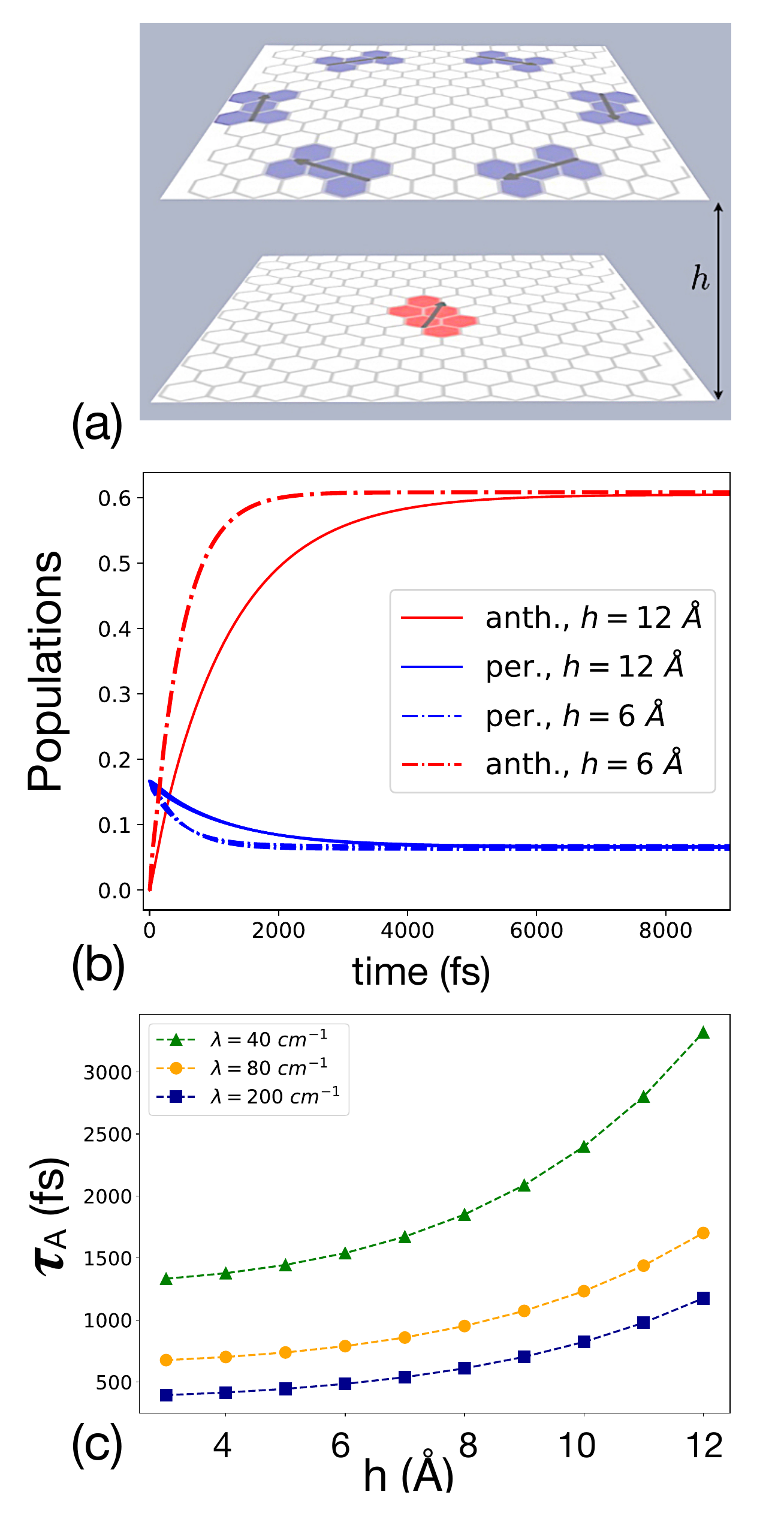}
  \caption{\label{fig7}(a) Two-layer antenna composed of symmetric perylene aggregate on top, and one anthanthrene on the bottom FG lattice. $h$ is the distance between layers. (b) Energy funnel to the acceptor: population of monomers for $h=6 \ \text{and} \ 12$ \r{A} taking $\lambda=200 \ \text{cm}^{-1}$ and $T=300$~K. (c) Transfer times obtained for different values of $h$ and $\lambda$ and $T=300$~K.}
\end{figure}
%
%
%

%
%
\begin{figure}
  \includegraphics[width=0.36\textwidth]{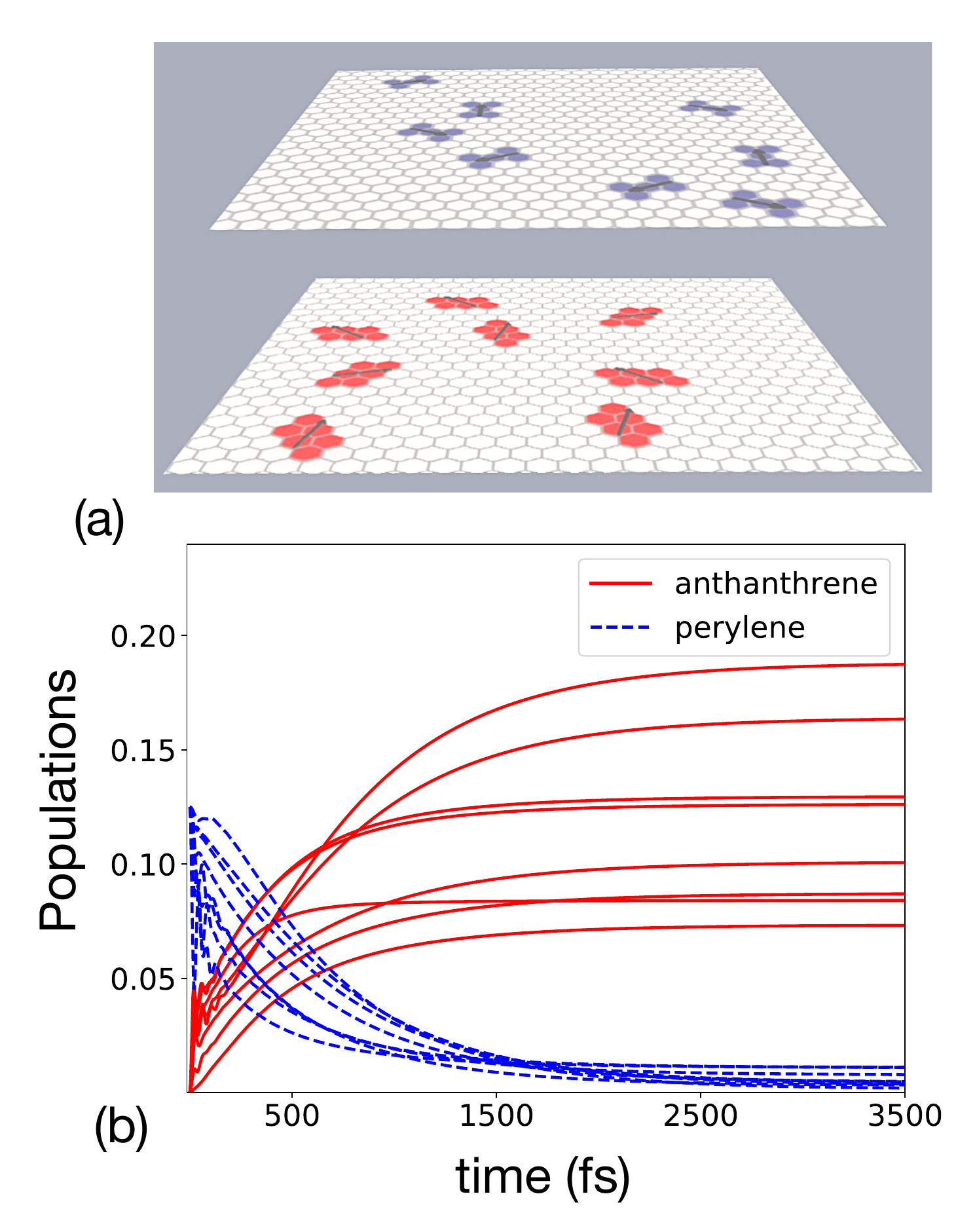}
  \caption{\label{fig8}(a) Larger two-layer antenna made of 16 randomly distributed perylene and anthanthrene moleculoids on separate sheets, with $h=6$ \r{A}. (b) Energy funnel of the antenna for $\lambda=80 \ \text{cm}^{-1}$ and $T=300$~K.}
\end{figure}

So far we have only discussed the situation with all moleculoids present on a single sheet of FG. In general it is conceivable that differently processed sheets, with different types of aggregate can be placed on one another, either directly, or using some spacers. In this situation, in order to concentrate excitation energy, we need to transfer it from one sheet to another. In this section, we consider such energy  funnelling in 3D situations. We consider the following two-layer examples: (1) a simple 7-site symmetric aggregate made of an all-perylene ring on one lattice together with an anthanthrene on a parallel FG sheet, Fig. \ref{fig7}a. And (2) a larger 16-site aggregate in which the two types of studied moleculoids are randomly distributed, each in its own FG lattices Fig. \ref{fig8}a. The 3D symmetric geometry can be compared to the 2D symmetric antenna, Fig. \ref{fig1}b, to also study the effect of the varying third dimension (inter-layer distance $h$) on the transfer time. We remind the reader that the 2D effective relative permittivity of FG was calculated $\epsilon^{eff}_r=0.7$ for the Coulombic interaction between moleculoids on the same lattice. Here, we made a conservative choice of inter-layer permittivity $\epsilon^{\perp}_r=1$ for the interaction of defects placed on different FG layers. The coupling between the sheets is therefore smaller than within the sheet, and consequently less favorable for energy transfer. Otherwise we assume the same environmental parameters for both FG lattices. We also take $T=300 \ K$ and $\gamma^{-1}=60$ fs for both two-layer simulations. Fig. \ref{fig7}b shows energy funnelling to the acceptor for two values of $h$ in the 3D symmetric aggregate. Once again the time-evolution of acceptor population can be well approximated by an exponential curve, leading to Fig. \ref{fig7}c which shows the transfer times for different inter-layer distances and reorganization energies. Typical distance between layers of FG has been reported around 6 \r{A} \cite{zbovril2010graphene, han2010unraveling}. Here, we take $h$ as a variable so that the results are applicable to other conditions, e.g. for distances created by putting spacers between the sheets. Distances smaller than 6 \r{A} may be considered as simulating any hypothetical larger values of couplings than those with $\epsilon^{eff}_r=1.0$. 

The second example concerns a larger area antenna cell as well as more moleculoids. Eight perylene-like and eight anthanthrene-like moleculoids are assumed to be randomly distributed in the top and bottom FG layers respectively. With larger number of moleculoids, the numerically exact HEOM calculations for the evolution of density matrix become extensively heavy. We consider 16 moleculoids for our largest sample in this study and take specific inter-layer distance $h=6$ \r{A} and reorganization energy $\lambda=80 \ \text{cm}^{-1}$ to show the evolution of excitations as they reach the acceptor sites. Due to the absence of symmetry in this sample, transfer times vary for different moleculoids, Fig. \ref{fig8}b. The exciton transfer to the anthanthrene acceptors range between 168 to 718 fs for the particular environmental parameters considered. The fast transfers in fact benefit from larger exciton delocalization over the antenna which is created by larger number of interacting moleculoids. This example clearly shows that although in larger antennae with larger inter-moleculoids distances the resonance couplings are generally weaker, increasing the number of interacting moleculoids will compensate for the low exciton delocalization caused by small inter-moleculoid $J_{ij}$'s. As a result, fast energy funnelling occurs both among defects arranged in small one-layer antenna cells and the large multi-layer sheets furnished with large enough number of defects.

\section{Conclusions}
\label{concl}
The molecular approach to the study of excitation energy transfer on Fluorographene lattice with graphene-like defects enables us in principle to examine arbitrarily large and random configurations of defects functioning as light-harvesting antennae. Standard methods of open quantum systems theory, as practised in the investigations of naturally occurring light-harvesting systems, may be used to such end. In this work, we have opted for exact method for propagation of excited states, to avoid uncertainties related to the validity of different perturbative methods. We studied different arrangements of selected defects (artificial light-harvesting antennae) on Fluorographene sheets subject to environment with different coupling strengths to the antenna electronic states.  While the crucial ingredient for energy funnelling on such antennae seems be the energy gap between the transition energies of different moleculoid types, we have also seen significant contribution from the geometrical parameters (orientation and position) of the defects composing the antennae. Extensive exciton delocalization contributing to level repulsion and decrease of the donor(antenna)--acceptor energy gap, is an important factor leading to fast downhill energy transfer accompanied by the corresponding spatial energy transfer within the Fluorographene sheet or between the sheets. Study of an ensemble of randomly organized antennae as well as a specific symmetric geometry, shows that for large reorganization energies, the energy transfer is as fast as in few hundred fs, and there is not much need to optimize the geometric structure of the antennae. This is particularly interesting for practical uses when one can only expect a limited control over structure and location of defects on the lattice. This study, although limited to a small antenna cell area, has examined the main electronic, geometric and environmental parameters involved in the excitation energy transfer. The fact that bath reorganization energy is found to have the most significant positive effect on the desired function of the proposed artificial antenna suggests that rather than targeting special precise geometric arrangements, engineering the spectral density of the environment, the Fluorographene sheet itself, is a viable route towards realizing artificial antenna based on defects in Fluorographene. 

To evaluate the prospect of the fluorographene-based biomimetic light-harvesting antennae in more detail, further theoretical work should consider defects of larger sizes and a greater variety of shapes. Also, methodology for calculating excited stated dynamics of extended Frenkel exciton systems composed of many sub-systems of the size treated in this work by cheap standard methods, parameterized by exact propagation methods, needs to be further developed. We will report on such progress elsewhere.

\begin{acknowledgments}
This work has been supported by Charles University Research Centre program No. UNCE/SCI/010, by the Neuron Fund for Support of Science through the Impuls 2014 grant, and by the Czech Science Foundation (GACR) through grant no. 18-18022S.
\end{acknowledgments}

\bibliography{StatTau}
\end{document}


\thispagestyle{empty}
\begin{center}
\Large{{\bf Energy Transfer Times in Fluorographene-based Biomimetic Light Harvesting Antennae: Supplementary Material}} \\
\vspace{8mm}
\large\text{Sayeh Rajabi and Tom\'{a}\v{s} Man\v{c}al}\footnote{mancal@karlov.mff.cuni.cz} \\
\vspace{5mm}
\normalsize{Faculty of Mathematics and Physics, Charles University,}\\ \normalsize{Ke Karlovu 5, 121 16 Prague 2, Czech Republic}
\end{center}

\section{Parameter space of random antennae}
\begin{figure}[h]
  \includegraphics[width=1\columnwidth]{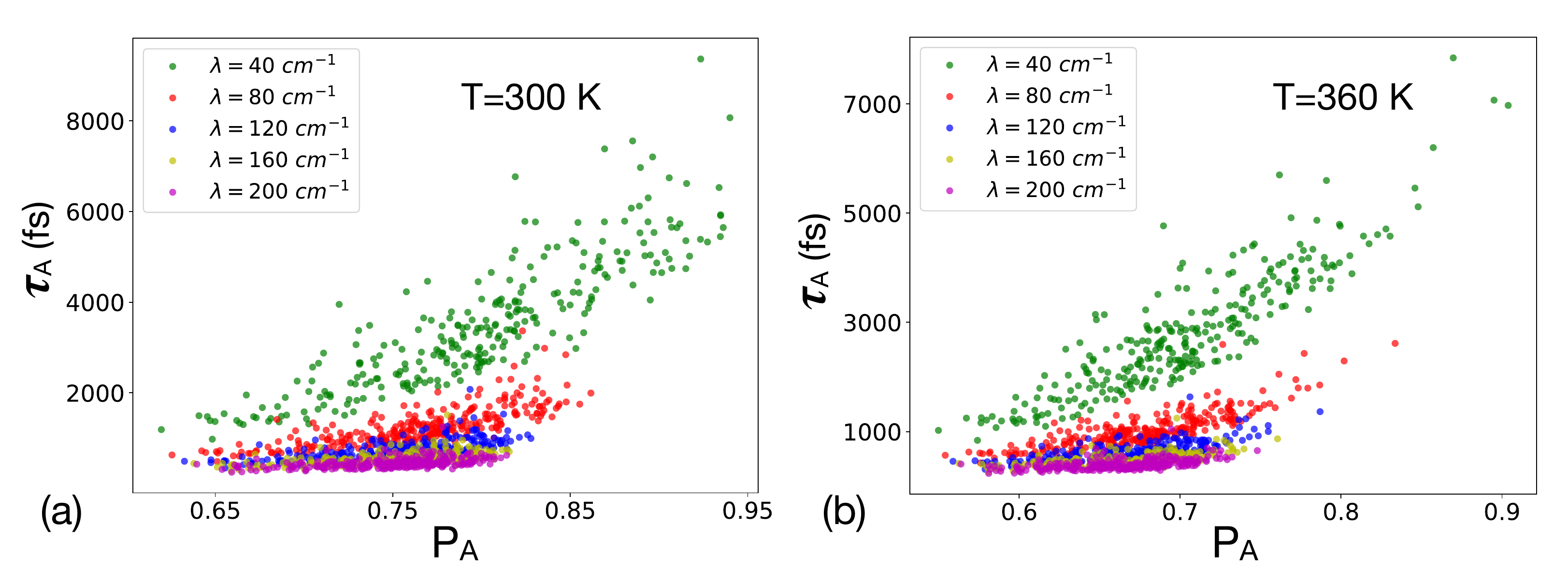}
  \caption{\label{fig1_SI}\small Transfer time and final population on acceptors in 280 random samples, with $\gamma^{-1}=60$ fs.}
\end{figure}

The parameter space of 280 random samples, $\tau_A$ vs $P_A$, is plotted in Fig. \ref{fig1_SI} showing the extension of values of the two antenna parameters for both temperatures considered in this study, $T=300$ and 360 K as well as different values of reorganization energy, $\lambda =$40, 80, 120, 160, 200 $\text{cm}^{-1}$. As was explained in the main text, excitation transport occurs faster when the reorganization energy increases. Therefore, geometric optimization mainly matters when the reorganization energy is small. 

\section{Effect of reorganization energy}
\begin{figure}
  \centering
  \includegraphics[width=1\textwidth]{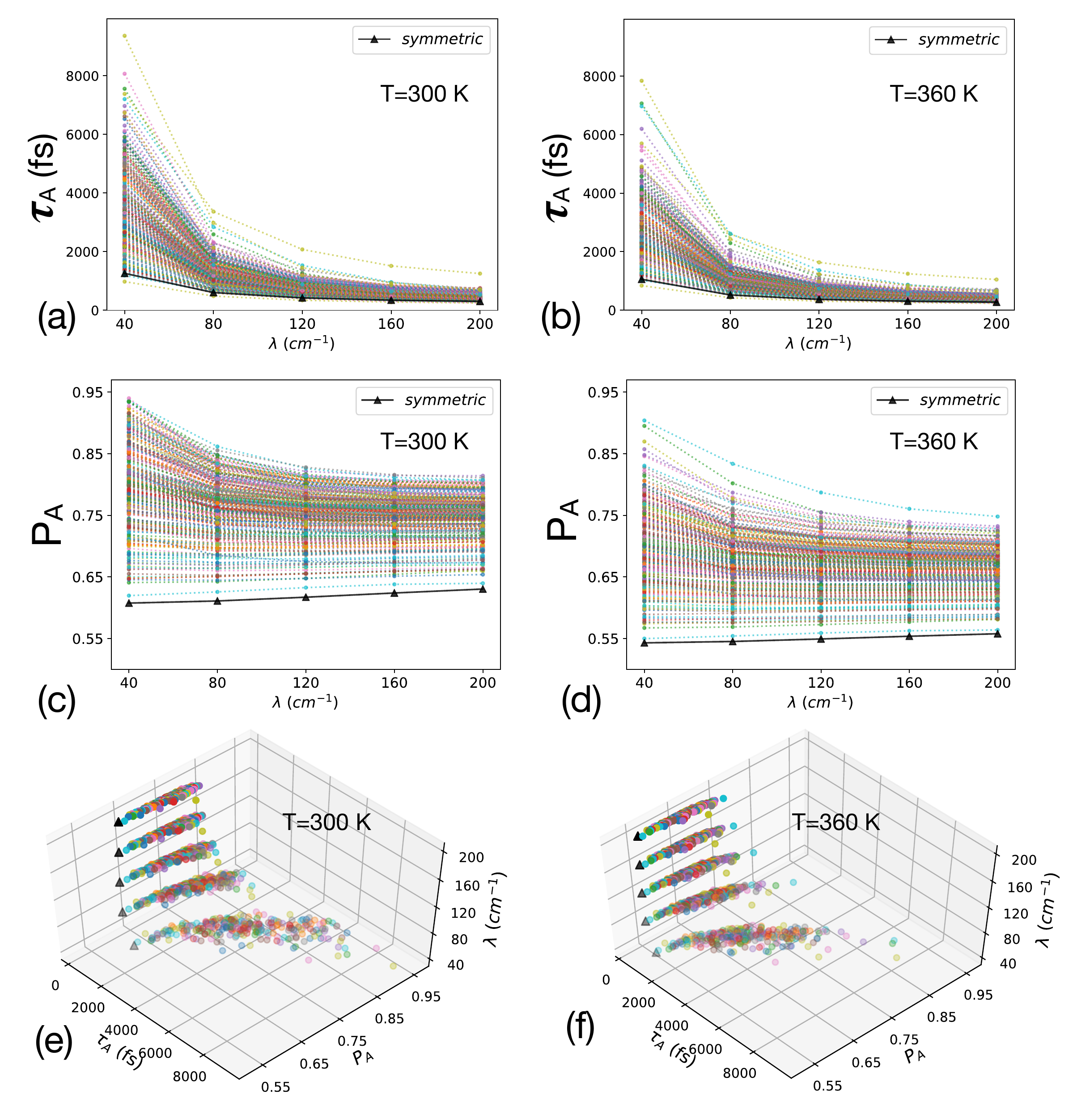}
  \caption{\label{fig2_SI}\small Transfer time and final population of acceptor as a function of the reorganization energy in the random samples, compared to the symmetric antenna, keeping the bath correlation time fixed at $\gamma^{-1}=60$ fs. Symmetric antenna results are depicted by triangles, all other antennae by circles.}
\end{figure}

Fig. \ref{fig2_SI} provides details of the dependence of $\tau_A$ and $P_A$ on the reorganizatin energy for both temperatures. Particularly, the change of final acceptor population with reorganization energy is negligible. Also, one can realize that the symmetric antenna is one of the fastest antennae for all values of $\lambda$ studied in this work.
%
%

\section{Variations of positions in the symmetric antenna}
\begin{figure}
  \centering
  \includegraphics[width=0.8\textwidth]{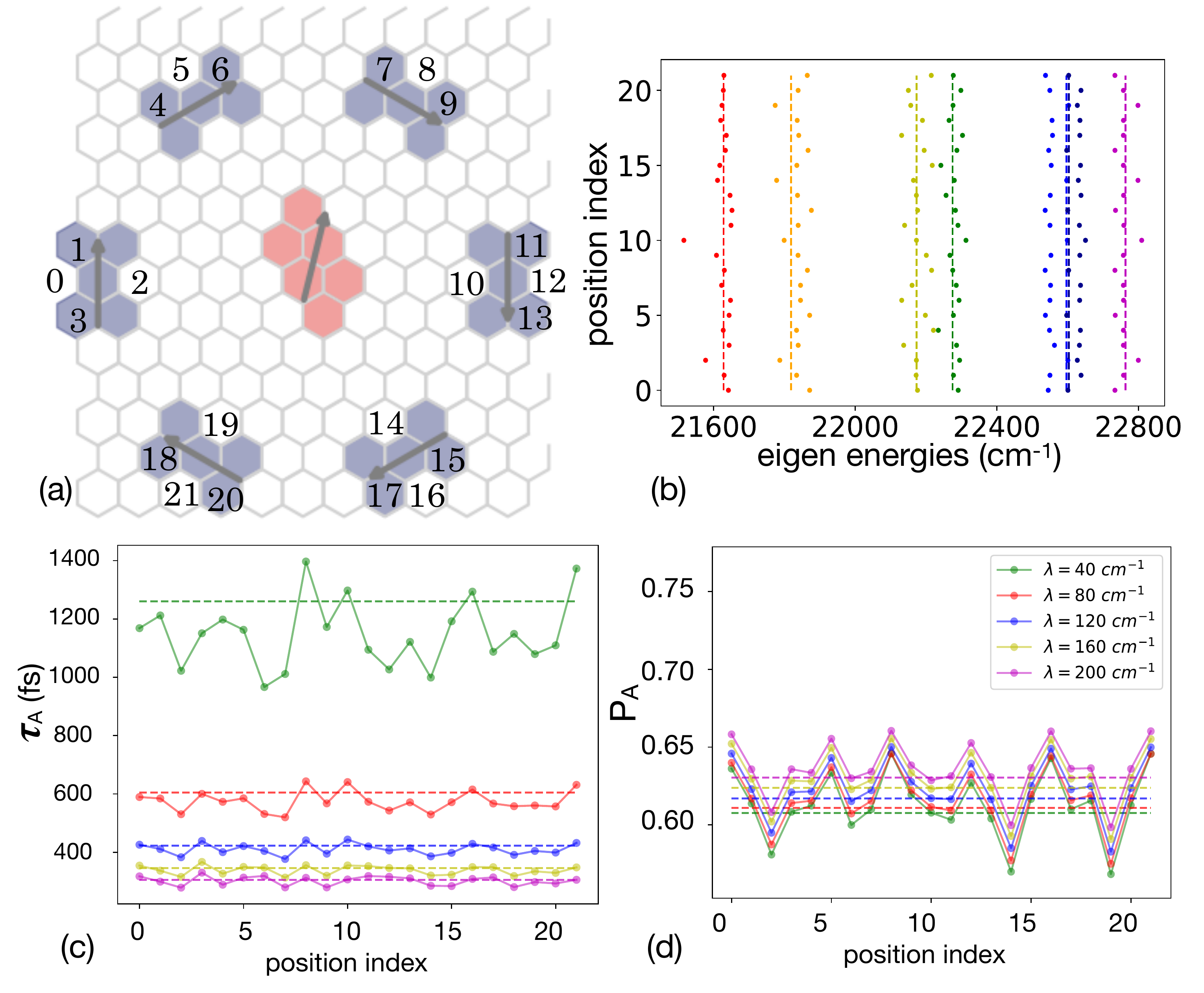}
  \caption{\label{fig3_SI}\small Symmetric antenna under variations of location, one moleculoid at a time. Numbers in (a)---position index in (b), (c) and (d)---represent the position of the center of the nearby moleculoid when relocated. (b) Distribution of eigen-energies of corresponding system Hamiltonians. (c) Transfer time and (d) acceptor's final population of the deformed antennae. Dashed lines in (b), (c) and (d) correspond to the values for the symmetric geometry.}
\end{figure}

Similar to the orientational variations around the symmetric geometry studied in the main text, we can study the robustness of this antenna in response to the changes in the location of defects. Fig. \ref{fig3_SI} summarizes the effect of one-lattice-site change in the position of each defect on the transfer time and acceptor population for various reorganization energies, fixed temperature $T=300$ K and fixed correlation time $\gamma^{-1}=60$ fs.
%
\section{Effect of correlation time}
\begin{figure}
  \centering
  \includegraphics[width=0.75\textwidth]{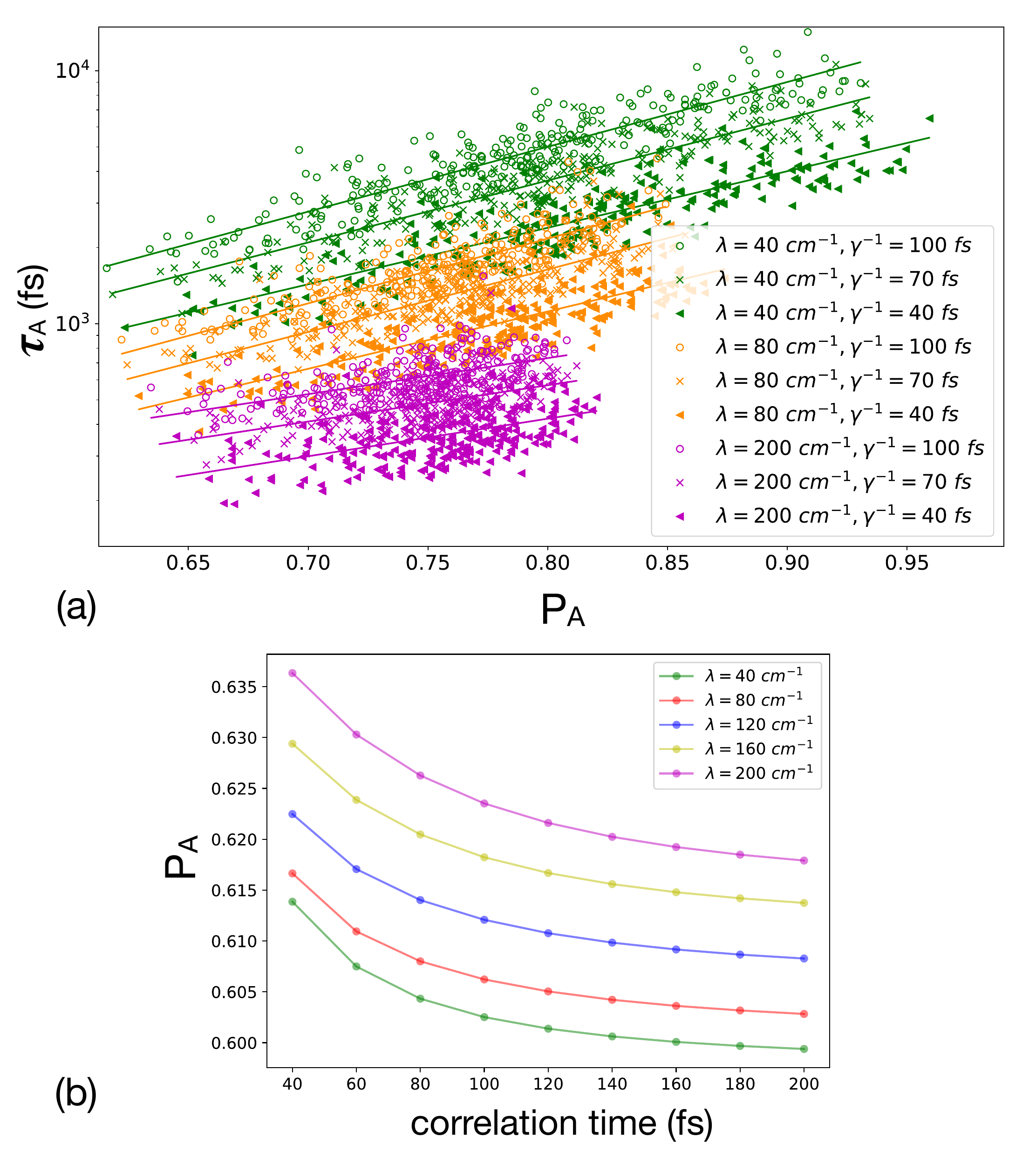}
  \caption{\label{fig4_SI}(a) Parameter space of 280 random samples for different values of reorganization energy and correlation time.  (b) Final population on the acceptor in the symmetric antenna for different values of correlation time.}
\end{figure}

Bath correlation time is another environmental parameter that affects the performance of an antenna. In Fig. \ref{fig4_SI}a, three values for correlation time and three for the reorganization energy present the parameter space of the random antennae. Smaller correlation time results in faster excitation transfer for all reorganization energies. $P_A$ insignificantly (and linearly) reduces as the correlation time increases, Fig. \ref{fig4_SI}b.
%
%


%